\documentclass[namedreferences]{solarphysics}

\usepackage[hyperref,optionalrh,showbiblabels]{spr-sola-addons} % For Solar Physics
\usepackage{graphicx}        % For eps figures, newer & more powerfull
\usepackage{color}           % For color text: \color command
\usepackage{breakurl}        % For breaking URLs easily trough lines
\def \arcsec {$^{\prime\prime}$}
\newcommand{\solphys}{{\it Solar Phys.}}

\chardef\us=`\_
\begin{document}

\begin{article}
\begin{opening}

\title{EUV and Magnetic Activities Associated with Type-I Solar Radio Bursts}

\author[addressref=aff1,email={licy001@163.com}]{\inits{C.Y.}\fnm{C.Y.}~\lnm{Li}}%\sep
\author[addressref=aff1,corref,email={yaochen@sdu.edu.cn}]{\inits{Y.}\fnm{Y.}~\lnm{Chen}}%\sep
\author[addressref=aff1]{\fnm{B.}~\lnm{Wang}}%\sep
\author[addressref={aff1,aff2}]{\fnm{G.P.}~\lnm{Ruan}}%\sep
\author[addressref=aff1]{\fnm{S.W.}~\lnm{Feng}}%\sep
\author[addressref=aff1]{\fnm{G.H.}~\lnm{Du}}%\sep
\author[addressref=aff1]{\fnm{X.L.}~\lnm{Kong}}%\sep
%\author[corref,email={e-mail.c@mail.com}]{\inits{S.}\fnm{Second}~\lnm{Author-c}}%\sep
%\author{\inits{T.}\fnm{Third}~\lnm{Author-x}}
%\author{\inits{}\fnm{}~\lnm{}\orcid{}}
%\author{P.~\surname{Author-a}$^{1}$\sep
%        E.~\surname{Author-b}$^{1}$\sep
%        M.~\surname{Author-c}$^{2}$
%       }

%   \institute{$^{1}$ First affiliation
%                     email: \url{e.mail-a} email: \url{e.mail-b}\\
%              $^{2}$ Second affiliation
%                     email: \url{e.mail-c} \\
%             }
\address[id=aff1]{Shandong Provincial Key Laboratory of Optical Astronomy and Solar-Terrestrial Environment, and Institute of Space Sciences, Shandong University, Weihai, Shandong 264209, China}
\address[id=aff2]{Key Laboratory of Solar Activity, National Astronomical Observatories, Chinese Academy of Sciences, Beijing 100012, China}
%\address[id=aff2]{Second affiliation}
%\address[id=aff3]{Third affiliation}

\runningauthor{C.Y.\ Li \emph{et al.}}
\runningtitle{Coronal Activities Associated with Type-I Bursts}

\begin{abstract}
Type-I bursts (\emph{i.e.} noise storms) are the earliest-known type of solar radio emission at the metre wavelength. They are believed to be excited by non-thermal energetic electrons accelerated in the corona. The underlying dynamic process and exact emission mechanism still remain unresolved. Here, with a combined analysis of extreme ultraviolet (EUV), radio and photospheric magnetic field data of unprecedented quality recorded during a type-I storm on 30 July 2011, we identify a good correlation between the radio bursts and the co-spatial EUV and magnetic activities. The EUV activities manifest themselves as three major brightening stripes above a region adjacent to a compact sunspot, while the magnetic field there presents multiple moving magnetic features (MMFs) with persistent coalescence or cancelation and a morphologically similar three-part distribution. We find that the type-I intensities are correlated with those of the EUV emissions at various wavelengths with a correlation coefficient of 0.7\,--\,0.8. In addition, in the region between the brightening EUV stripes and the radio sources there appear consistent dynamic motions with a series of bi-directional flows, suggesting ongoing small-scale reconnection there. Mainly based on the induced connection between the magnetic motion at the photosphere and the EUV and radio activities in the corona, we suggest that the observed type-I noise storms and the EUV brightening activities are the consequence of small-scale magnetic reconnection driven by MMFs. This is in support of the original proposal made by Bentely \emph{et al.} (\solphys\ 193, 227, 2000).
\end{abstract}
\end{opening}
%-------------------------------------------------
\section{Introduction}
     \label{S-Introduction}
Type-I solar radio bursts, also known as noise storms, represent one of the earliest-discovered solar radio bursts at metre wavelengths (\citealp{Hey46}; \citealp{Elgaroy77}). These bursts are in general believed to be excited by nonthermal energetic electrons accelerated in the corona, with a long duration sometime lasting for days to a week. Type-I solar radio bursts usually occur above large sunspots, with polarization consistent with o-mode electromagnetic waves propagating in plasmas (\emph{e.g.} \citealp{Mercier84}; \citealp{Kai85}). The emission consists of two components: one is a long-lasting wide-band continuum component with relatively weak brightness; the other is a very narrow band component of high brightness temperature ($T_\mathrm{B}$). The exact emission mechanism and dynamic process accounting for type-I radio bursts still remain elusive. Often type-I noise storms are observed during quiet (non-flaring) times, and their duration is usually longer than that of a solar flare, so it is not necessary for a type-I burst to be directly related to a flare. Yet, there exists evidence that type-I bursts are modulated or even generated by coronal mass ejections (CMEs) (\emph{e.g.} \citealp{cher01}; \citealp{Kath07}; \citealp{iwai12a}). Our study mainly concerns type-I bursts occurring during non-flaring times.

Different scenarios have been proposed for the underlying dynamic process, with the general belief that type-I bursts are likely associated with the gradual restructuring of the coronal magnetic field. The magnetic restructuring is either driven by photospheric magnetic activities such as flux emergence (\emph{e.g.} \citealp{Zwaan78}; \citealp{Zwaan85}) or moving magnetic features (MMFs: \citealp{Sheeley69}; \citealp{Harvey73}; \citealp{Hagenaar05}; \citealp{Bentley00}), or corresponds to so-called interchange reconnection between closed and open magnetic fields (\citealp{Crooker02}; \citealp{Fisk05}; \citealp{Zanna11}; \citealp{Edmondson12}).

In the flux emergence scenario, there exist two assumptions for the acceleration of type-I emitting energetic electrons, either by reconnection between the newly emerged flux and the ambient coronal magnetic field (\citealp{Benz81}), or through the generation of lower hybrid waves at the interface of the emerging (and expanding) flux and the coronal background (\citealp{Spicer82}). Observations supporting this scenario show that the type-I bursts took place after a significant flux emergence was observed, indicating a possible cause-effect relationship. Yet, there also exist observations indicating that flux emergence is not necessary for a noise storm to occur \citep{Bentley00,Willson05a,Willson05b}.

\citet{Bentley00} proposed that ``MMFs are at the origin of the observed metric noise storms''. MMFs, first reported by \citet{Sheeley69}, are magnetic poles (or bipoles) moving outward along the radial direction from a sunspot, with a velocity of 0.1\,--\,2 km s$^{-1}$. In the course of the MMF motion, magnetic cancellation and merging are frequently observed, sometimes associated with X-ray transient brightenings (\citealp{Shimizu94}), and surge and X-ray jet activity (\citealp{Canfield96}), indicating the occurrence of magnetic reconnection. \citet{Bentley00} therefore suggested that the reconnection driven by MMFs, when approaching other magnetic features (\emph{e.g.} the moat boundary of the spot) of opposite polarity, produces the type-I emitting energetic electrons.

The interchange reconnection scenario for type-I radio bursts was proposed by \citet{Zanna11}, working as a unified interpretation for persistent coronal outflows, likely the source of the solar wind from the boundary of the active region, and type-I radio bursts. Their suggestion was based on the correlation analysis of Doppler shift measurements in the relevant part of the active region, the radio sources given by the \emph{Nan\c{c}ay Radioheliograh} (NRH: \citealp{Kerdraon97}) as well as the coronal magnetic field configuration deduced with an extrapolation method.

As seen from the above review of earlier studies, there exists a long-standing interest in finding the coronal and photospheric counterparts of the type-I burst. Yet, no consensus has been reached regarding the relationship of type-I bursts with the phenomena observed in the corona (in EUVs or X-rays) and at the photosphere (in the magnetic field). For example, \citet{iwai12b} found that the causal relationship between the observed soft X-rays (SXR) activity and type-I onset is unclear, while \citet{Willson05a} could not identify correlated magnetic activity using the \emph{Michelson Doppler Imager} (MDI) on board the \emph{Solar and Heliospheric Observatory} (SOHO) spacecraft. In another separate study, \citet{Willson05b} concluded that magnetic activities were found 6h before (rather than during) a type-I burst. In \citet{Bentley00} with the MMF scenario, no correlated time evolutionary characteristics between the radio emission and the magnetic activities were revealed. This is likely due to the limited quality of earlier data collected by the MDI/SOHO for the photospheric magnetic field and the \emph{Transition Region and Coronal Explorer} (TRACE) for the coronal EUV activity. Note that \citet{Bentley00} did report a transient EUV bright point and occasional loop brightenings, however, their connection to the long-term persistent type-I bursts has not been clearly established.

The above non-satisfactory state of relevant studies reveals a strong need to re-explore the origin of noise storms with the latest-available high quality data, especially those with unprecedentedly high resolution at multi EUV wavelengths recorded by the \emph{Atmospheric Imaging Assembly} (AIA: \citealp{Lemen12}) as well as vector magnetic field measurements by the \emph{Helioseismic and Magnetic Imager} (HMI: \citealp{Schou12}) (with a temporal resolution of 45s, in comparison with the 96 min of MDI data) on board the \emph{Solar Dynamics Observatory} (SDO: \citealp{Pesnell12}). In order to identify the presence and location of the noise storm, it is necessary to conduct a combined analysis of these data with the multi-frequency radio imaging data of NRH. The purpose of this article is to report magnetic activities in the corona and at the photosphere that seem to be strongly associated with a type-I radio burst, on the basis of a case study using these high-quality data observed at various wavelengths.

\section{Observations and Event Overview}
      \label{S-general}
According to the dynamic spectral data provided by the \emph{Compact Astronomical Low-cost Low-frequency Instrument for Spectroscopy and Transportable Observatory} (CALLISTO-BLEN) radio spectrograph (170\,--\,870 MHz with a cadence of 0.25s), the type-I radio burst of study appears at 15:00 UT on 29 July 2011, and lasts for nearly a week till the radio sources are rotated to the backside of the Sun as seen from the NRH images. We also refer to the spectral data of the ARTEMIS IV Multichannel Radiospectrograph (20\,--\,650 MHz) with a cadence of 0.1s. The spectra show that the type-I radio bursts mainly occurs between 150\,--\,300 MHz (see Figure 1 for the radio spectra and sources observed on 30 July 2011). The radio imaging data of NRH most relevant to our study are at 298, 270, and 228 MHz. These data are used to determine the location and flux density (and the $T_\mathrm{B}$) of the emission source.

In Figure 1d, we show the temporal profiles of $T_\mathrm{B}$ maxima and polarization levels observed by NRH from 09:00 to 15:00 UT on 30 July. We see that the $T_\mathrm{B}$ maxima vary in the range of $10^7$ to $10^9$ K and the emission is strongly polarized in the left-handed sense at a level of $>$ 80\,\%. On this day, the radio source is located close to the solar meridian. This brings a relatively more accurate measurement of the magnetic field at the photosphere, with less projection effect of the foreground coronal loops on the EUV activity at lower altitude that may be of interest. Therefore, in this study we focus on the type-I burst observed on this day.

The magnetograms (cadence: 45s; pixel size: 0.6\arcsec) provided by HMI and the EUV data provided by AIA (cadence: 12s; pixel size: 0.6\arcsec) at different passbands (corresponding to different temperatures) are examined to explore the accompanying activities in the corona. The SXR images recorded by the \emph{Soft X-ray Imager} on board the \emph{Geostationary Operational Environmental Satellite} (SXI/GOES-15) (cadence: 1 min; pixel size: 5\arcsec) are analyzed to determine whether accompanying X-ray activity exists, while no data from the \emph{X-Ray Telescope} (XRT: \citealp{Golub07}) on board the \emph{Hinode} (Solar B: \citealp{Kosugi07}) (XRT/\emph{Hinode}) are available during the interval of interest.

In Figure 2a\,--\,2b, we present the HMI continuum intensity map and the magnetogram superposed with the contours of the NRH radio brightness obtained at about the same time. The figure illustrates the relative location of the radio sources and the leading spot of the active region (NOAA AR 11260). As told from the long-term evolution of the NRH and SDO data, while rotating together, the radio sources are located ahead of and at the northwestern direction with respect to the leading spot. This spatial configuration indicates that the type-I burst is associated with the magnetic evolution of the leading spot. According to the online Coordinated Data Analysis Workshop (CDAW) catalogue, only one narrow jet-like CME with an angular width of $\approx$7 deg and a linear speed of 304 km s$^{-1}$ was detected at about 07:48 UT by the \emph{Large Angle and Spectrometric Coronagraph} (LASCO, \citealp{Bru95}), around 10 hours preceding the start of the type-I burst of study. No observable eruptive signatures are associated with this narrow CME, according to the AIA data. In addition, no solar flares are released from the AR 11260 during the time of interest. This indicates that the type-I burst investigated here is not associated with solar eruptive phenomena. Therefore, as in most cases reported earlier, the energetic electrons emitting type-I radio bursts are accelerated by non-flaring processes.

The AR first appears in the SDO field-of-view (FOV) on 24 July 2011, with rapid evolution caused by significant flux emergence and merging of magnetic structures. On 29 July 2011, the AR forms a major leading spot with a negative polarity. The spot maintains its dominance of the AR magnetic configuration until a new large positive polarity appears along its western side on 3 August. During this stage of the AR evolution, around the leading spot there appear persistent unipolar or bipolar magnetic features of much smaller scale moving along an almost radial direction. These polarities are called MMFs, as mentioned earlier.

Figure 2c shows the coronal loop system associated with the AR. A majority of the loops from the leading spot connect to the positive polarity of the AR. Along the direction to the radio source, a set of loops depicts the northwestward expansion of the AR, consistent with the above conjecture that the radio burst is associated with the leading spot. In the following section, we conduct a detailed analysis to further establish the correlation between the magnetic and EUV activities.

\section{Magnetic and EUV Activities and their Correlation with the Type-I Radio Burst on 30 July 2011}
As seen from Figure 3 and the accompanying movie, MMFs within the northwest neighborhood of the leading spot are very active on 30 July 2011. Some positive polarities associated with the MMFs stop their motion at a distance of about 20\arcsec\ away from the nearest outer border of the leading spot, and merge there to form a wall-like structure. The wall-feature is clearly observed from 09:00 to 13:00 UT. Later, the structure disintegrates into several patches of smaller size. The boundary of the leading spot and the positive wall-like structure together form an overall bipolar regime within which there exist several non-moving magnetic patches of negative polarity. During most of the interval covered by Figure 3, these internal patches can be grouped into three parts (see green arrows).

During the type-I burst on 30 July, the MMFs continue to separate from the leading spot and move towards the wall-like structure. During this motion, the MMFs may merge or cancel with pre-existing magnetic polarities. In particular, the set of three pre-existing negative polarities as well as the border serve as major regions of magnetic merging and cancellation (best viewed from the accompanying movie).

In Figures 2c\,--\,2f, we show the AIA image at 171 \AA　 within a larger FOV and the images at 304, 211, 131 \AA　 within a smaller FOV. We see that underneath the loops that extend from the leading spot towards the radio sources, there exist several sets of small-scale brightening point-like or stripe-like structures. As seen from the accompanying movie, the associated EUV activity becomes more obvious after 09:33 UT. From 10:40 to 12:30 UT, the brightening structures can be approximately separated into three stripes. Note that although these structures change constantly in brightness, their relative locations seem to be relatively fixed (see the white arrows in Figures 2 and 3). The stripes are 2\,--\,5\arcsec\ wide and 10\,--\,15\arcsec\ long. In the image at 304 \AA　 of cooler temperature, the 171 \AA　 loops vanish due to their higher temperature (and higher altitudes). The brightening is observed through most AIA EUV passbands with a very similar morphology. This indicates that the emitting source consists of plasmas with multi-temperatures or with a broad velocity distribution function.

To reveal the correlation between the activities observed by HMI-AIA and those observed in the radio metre wavelengths by NRH, we select a box to cover the brightening region. We first sum up the photons along the radial direction pointing away from the leading spot. The obtained temporal profiles of photon numbers thus represent the EUV emission distribution along the other direction. The profiles are stacked together to make a time-distance map as shown in Figure 4 for the data at 211 \AA. The clear three-stripe pattern is very obvious as seen from this figure. At a different location, the EUV brightening starts at a slightly different time. Note that we point out that from 09:40 to 14:50 UT the HMI magnetic field also presents a three-part distribution of negative polarity. This magnetic distribution seems to strongly affect the site of magnetic cancelation. In addition, comparing the magnetic and EUV data (Figure 3), we find that the EUV brightening stripes are co-spatial with the specific pattern of magnetic distribution. This indicates that the EUV activity is driven by the underlying MMF motion.

We then sum up all photons within the box, and this gives curves that are proxies of the EUV emission intensity, as shown in Figure 4b for the 211 \AA　 data as an example (in green). The overlaid curves give the $T_\mathrm{B}$ maxima obtained by NRH at the three frequencies 298 (black), 270 (red), and 228 (blue) MHz. Overall a nice correlation is seen between the different sets of curves, as further supported by our quantitative correlation analysis that reveals high correlation coefficients of 0.7\,--\,0.8 between the 211 \AA　 summed photon number curve and the $T_\mathrm{B}$ maxima at different frequencies. Similar high correlations are found for other AIA EUV passbands such as 131, 94, 335 \AA.

The above analysis presents major evidence supporting the original proposal by \citet{Bentley00} that MMFs are the origin of type-I radio bursts. Here we establish this cause-effect relationship by connecting the radio emission to the EUV activity, and the EUV activity to the MMF activity. It has been suggested by \citet{Bentley00} that the MMF dynamics may trigger reconnection in the corona. The magnetic cancellation associated with the MMFs provides evidence for this suggestion. Further evidence of reconnection can be obtained from the following two aspects of the EUV data.

Firstly, we observe several bi-directional plasma flows during the process. The flows start from a common site close to, yet higher than, the brightening stripes, and move towards and away from the site. This is demonstrated in Figure 5 and the accompanying movie (see the arrows). The separation speed of the bidirectional flows is about 35 km s$^{-1}$ and 51 km s$^{-1}$ for the two cases. Secondly, there appear considerable perturbations in the EUV images along the direction pointing away from the leading spot to the radio sources (see the AIA movie accompanying Figure 2). These outer EUV perturbations, together with the inner brightening structures, are likely associated with the two necessary components of reconnection directed towards and away from the presumed reconnection site. Reconnection driven by the MMFs may take place between the two regions and the energetic electrons released may move to the larger outer magnetic structures and excite a type-I radio burst there.

To show this picture more clearly, we conduct a potential-field source-surface (PFSS) extrapolation (\citealp{Schat69}; \citealp{Schri03}) of the large-scale magnetic field configuration in the corona. The result, based on two successive Carrington maps (CR 2112\,--\,2113), is shown in Figure 6a. Along the direction of interest, the field lines expand from the leading spot towards the northwest. This is consistent with the inclined loops observed by AIA at 171 \AA. The loop inclination is along the direction pointing from the MMFs (and the corresponding EUV activity) at the photosphere to the radio sources relatively high in the corona. Thus, the deviation of radio sources from the MMFs (as seen from Figure 2) can be attributed to a projection effect.

In Figure 6b, we present a schematic illustrating our picture of the generation of the type-I burst. The complex quadrupolar configuration is given by the alignment of the wall-like positive structures, the pre-existing negative polarities, and the positive part of the MMFs, and the negative leading spot, as described earlier. The MMFs (which possibly connect back to the border of the leading spot) move towards the pre-existing negative polarities and lead to significant magnetic cancellation there. The moving MMFs drive the associated loops in the corona to interact with nearby loops (which connect the pre-existing negative polarities and the positive wall-like structures). This interaction leads to reconnection in the corona (as indicated by the star symbol in Figure 6b). The resultant bi-directional flows/perturbations lead to EUV activity at both higher and lower locations (see the notations in Figure 6b).

Electrons can be accelerated during the process and become type-I emitting. The radio sources are along the higher part of post-reconnection loops with roots in the positive polarity. Note that this MMF-driven reconnection picture for type-I bursts is very similar to what proposed initially by \citet{Bentley00}, except that in our events there exist significant positive polarities (the wall-like structures) ahead of the leading negative spot. This explains why the polarization of the radio bursts is left-handed while the leading spot is of negative polarity, with the general conclusion that the type-I burst is in the o-mode (see, \emph{e.g.}, \citealp{Mercier84}; \citealp{Kai85}). The energetic electrons moving towards the lower solar atmosphere along post-reconnection loops find a denser and more-collisional environment, where the conditions of type-I emission probably cannot be satisfied. This tentatively explains why the type-I bursts only appear above a certain frequency (say, 300 MHz in this case). Further investigations, such as a differential emission measure (DEM) analysis, will be required to confirm the presence of dense loops above the MMFs.

To answer the question of whether MMFs are related to an enhancement in X-rays, we examine the data recorded by the Solar X-ray Imager on GOES-15 (the data from XRT/\emph{Hinode} at higher resolution are not available as mentioned). Weak signatures of X-ray activity are indeed found in the region of enhanced EUV emissions, yet, due to the low resolution and sensitivity of the XRT data, the result is not conclusive. This leaves open the question of whether the type-I radio bursts have counterparts in X-ray.

\section{Summary and Discussion}
In this study, we present an analysis of a type-I solar radio burst combining the EUV data provided by AIA/SDO, magnetic field data observed at the photosphere by HMI/SDO, and radio imaging data from NRH as well as dynamic spectra from the Callisto-Blen and Artemis spectrographs. The main purpose is to determine the activity in the corona and at the photosphere associated with solar type-I radio bursts (or noise storms). It is found that the radio bursts are highly correlated with some small-scale EUV brightening stripes that are likely caused by MMFs flowing outside from the leading sunspot at the photosphere. This establishes the connection between photospheric magnetic activity, EUV activity, and the solar noise storm. The study supports the earlier proposal made by \citet{Bentley00} that MMFs can drive magnetic reconnection in the corona that further releases non-thermal energetic electrons to excite the radio burst. The presence of reconnection is supported by the observation of cancellation of MMFs with nearby magnetic features, brightening in the EUV passbands, and bidirectional plasma flows in the region of interest, as well as their correlation with the type-I radio burst.

In the Introduction, we present various scenarios proposed to explain the dynamical process accounting for the release of type-I emitting energetic electrons. In general, it is likely that different events occurring at various stages of ARs may have different driving mechanisms. For example, the type-I radio bursts are possibly associated with flux emergence and interchange reconnection during the developing stage of an AR, since during this stage the AR is characterized by rapid flux emergence. This may cause a significant expansion and therefore interaction of the AR with a nearby coronal hole if there is one. On the other hand, during the steadily-evolving or decay stage of an AR, the MMFs may become an important feature of magnetic activity surrounding the AR. Our case study here provides support to the MMF scenario. Regarding the other two scenarios, earlier observations suffer from low resolution photospheric magnetic field measurements and a low sensitivity as well as limited temperature coverage in the EUV (X-ray) data, and therefore the published observational evidence supporting them are not sufficiently robust and further evidence using the present high-quality data is needed. In addition, as mentioned earlier, CMEs may also play a role in modulating or even generating type-I radio bursts (see, \emph{e.g.}, \citealp{iwai12a}). More cases should be analyzed before we draw a firm conclusion regarding the exact role of different processes.

\begin{acks}
We thank the teams of e-Callisto, ARTEMIS, NRH, AIA/SDO, and HMI/SDO for making their data available to us. This work was supported by the National Natural Science Foundation of China (41331068, 11503014, U1431103 and U13311013), and the Key Laboratory of Solar Activity of the Chinese Academy of Sciences (CAS) under the grant number KLSA201602.
\end{acks}

\paragraph*{\footnotesize Disclosure of Potential Conflicts of Interest} \footnotesize The authors declare that they have no conflicts of interest.

%%% BIBLIOGRAPHY %%%%%%%%%%%%%%%%%%%%%%%%%%%%%%%%%%%%%%%%%%%%%%%%%%%%%%%%%%%

     % format of references provided by the journal (.bst)
\bibliographystyle{spr-mp-sola}
     % name your Bibtex file containing your references (.bib)
\bibliography{reference}

\begin{thebibliography}{33}
% BibTex style file: spr-mp-sola.bst (nameyear), 2015-03-09
\ifx\bisbn     \undefined \def\bisbn  #1{ISBN #1}\fi
\ifx\binits    \undefined \def\binits#1{#1}\fi
\ifx\bauthor   \undefined \def\bauthor#1{#1}\fi
\ifx\batitle   \undefined \def\batitle#1{#1}\fi
\ifx\bjtitle   \undefined \def\bjtitle#1{\textit{#1}}\fi
\ifx\bvolume   \undefined \def\bvolume#1{\textbf{#1}}\fi
\ifx\byear     \undefined \def\byear#1{#1}\fi
\ifx\bissue    \undefined \def\bissue#1{#1}\fi
\ifx\bfpage    \undefined \def\bfpage#1{#1}\fi
\ifx\blpage    \undefined \def\blpage #1{#1}\fi
\ifx\burl      \undefined \def\burl#1{\textsf{#1}}\fi
\ifx\href      \undefined \def\href#1#2{\textsf{#2}}\fi
\ifx\betal     \undefined \def\betal{\textit{et al.}}\fi
\ifx\bctitle   \undefined \def\bctitle#1{#1}\fi
\ifx\beditor   \undefined \def\beditor#1{#1}\fi
\ifx\bbtitle   \undefined \def\bbtitle#1{\textit{#1}}\fi
\ifx\bedition  \undefined \def\bedition#1{#1}\fi
\ifx\bseriesno \undefined \def\bseriesno#1{\textbf{#1}}\fi
\ifx\blocation \undefined \def\blocation#1{#1}\fi
\ifx\bsertitle \undefined \def\bsertitle#1{\textit{#1}}\fi
\ifx\bsnm      \undefined \def\bsnm#1{#1}\fi
\ifx\bsuffix   \undefined \def\bsuffix#1{#1}\fi
\ifx\bparticle \undefined \def\bparticle#1{#1}\fi
\ifx\barticle  \undefined \def\barticle#1{}\fi
\ifx\binstitute  \undefined \def\binstitute#1{#1}\fi
\ifx\bpublisher  \undefined \def\bpublisher#1{#1}\fi
\ifx\doiurl    \undefined
  \def\doiurl#1{\href{http://dx.doi.org/#1}{\textsf{DOI}}}\fi
\ifx\arxivurl  \undefined
  \def\arxivurl#1{\href{http://arxiv.org/abs/#1}{\textsf{arXiv}}}\fi
\ifx\adsurl    \undefined
  \def\adsurl#1{\href{http://adsabs.harvard.edu/abs/#1}{\textsf{ADS}}}\fi
\ifx\botherref \undefined \def\botherref#1{}\fi
\ifx\url       \undefined \def\url#1{\textsf{#1}}\fi
\ifx\bchapter  \undefined \def\bchapter#1{}\fi
\ifx\bbook     \undefined \def\bbook#1{}\fi
\ifx\bcomment  \undefined \def\bcomment#1{#1}\fi
\ifx\oauthor   \undefined \def\oauthor#1{#1}\fi
\ifx\citeauthoryear \undefined\def \citeauthoryear#1{#1}\fi
\ifx\endbibitem\undefined \def\endbibitem{}\fi
\ifx\bconflocation  \undefined \def\bconflocation#1{#1} \fi

\bibitem[\protect\citeauthoryear{{Bentley} \textit{et~al.}}{2000}]{Bentley00}
\begin{barticle}
\bauthor{\bsnm{{Bentley}}, \binits{R.D.}},
\bauthor{\bsnm{{Klein}}, \binits{K.-L.}},
\bauthor{\bsnm{{van Driel-Gesztelyi}}, \binits{L.}},
\bauthor{\bsnm{{D{\'e}moulin}}, \binits{P.}},
\bauthor{\bsnm{{Trottet}}, \binits{G.}},
\bauthor{\bsnm{{Tassetto}}, \binits{P.}},
\bauthor{\bsnm{{Marty}}, \binits{G.}}:
\byear{2000},
\batitle{{Magnetic Activity Associated With Radio Noise Storms}}.
\bjtitle{\solphys}
\bvolume{193},
\bfpage{227}.
\doiurl{10.1023/A:1005218007132}.
\adsurl{http://ads.bao.ac.cn/abs/2000SoPh..193..227B}.
\end{barticle}
\endbibitem

\bibitem[\protect\citeauthoryear{{Benz} and {Wentzel}}{1981}]{Benz81}
\begin{barticle}
\bauthor{\bsnm{{Benz}}, \binits{A.O.}},
\bauthor{\bsnm{{Wentzel}}, \binits{D.G.}}:
\byear{1981},
\batitle{{Coronal evolution and solar type I radio bursts - an ion-acoustic
  wave model}}.
\bjtitle{\aap}
\bvolume{94},
\bfpage{100}.
\adsurl{http://ads.bao.ac.cn/abs/1981A\%26A....94..100B}.
\end{barticle}
\endbibitem

\bibitem[\protect\citeauthoryear{{Brueckner} \textit{et~al.}}{1995}]{Bru95}
\begin{barticle}
\bauthor{\bsnm{{Brueckner}}, \binits{G.E.}},
\bauthor{\bsnm{{Howard}}, \binits{R.A.}},
\bauthor{\bsnm{{Koomen}}, \binits{M.J.}},
\bauthor{\bsnm{{Korendyke}}, \binits{C.M.}},
\bauthor{\bsnm{{Michels}}, \binits{D.J.}},
\bauthor{\bsnm{{Moses}}, \binits{J.D.}},
\bauthor{\bsnm{{Socker}}, \binits{D.G.}},
\bauthor{\bsnm{{Dere}}, \binits{K.P.}},
\bauthor{\bsnm{{Lamy}}, \binits{P.L.}},
\bauthor{\bsnm{{Llebaria}}, \binits{A.}},
\bauthor{\bsnm{{Bout}}, \binits{M.V.}},
\bauthor{\bsnm{{Schwenn}}, \binits{R.}},
\bauthor{\bsnm{{Simnett}}, \binits{G.M.}},
\bauthor{\bsnm{{Bedford}}, \binits{D.K.}},
\bauthor{\bsnm{{Eyles}}, \binits{C.J.}}:
\byear{1995},
\batitle{{The Large Angle Spectroscopic Coronagraph (LASCO)}}.
\bjtitle{\solphys}
\bvolume{162},
\bfpage{357}.
\doiurl{10.1007/BF00733434}.
\adsurl{http://ads.bao.ac.cn/abs/1995SoPh..162..357B}.
\end{barticle}
\endbibitem

\bibitem[\protect\citeauthoryear{{Canfield} \textit{et~al.}}{1996}]{Canfield96}
\begin{barticle}
\bauthor{\bsnm{{Canfield}}, \binits{R.C.}},
\bauthor{\bsnm{{Reardon}}, \binits{K.P.}},
\bauthor{\bsnm{{Leka}}, \binits{K.D.}},
\bauthor{\bsnm{{Shibata}}, \binits{K.}},
\bauthor{\bsnm{{Yokoyama}}, \binits{T.}},
\bauthor{\bsnm{{Shimojo}}, \binits{M.}}:
\byear{1996},
\batitle{{H alpha Surges and X-Ray Jets in AR 7260}}.
\bjtitle{\apj}
\bvolume{464},
\bfpage{1016}.
\doiurl{10.1086/177389}.
\adsurl{http://ads.bao.ac.cn/abs/1996ApJ...464.1016C}.
\end{barticle}
\endbibitem

\bibitem[\protect\citeauthoryear{{Chertok} \textit{et~al.}}{2001}]{cher01}
\begin{barticle}
\bauthor{\bsnm{{Chertok}}, \binits{I.M.}},
\bauthor{\bsnm{{Kahler}}, \binits{S.}},
\bauthor{\bsnm{{Aurass}}, \binits{H.}},
\bauthor{\bsnm{{Gnezdilov}}, \binits{A.A.}}:
\byear{2001},
\batitle{{Sharp Decreases of Solar Metric Radio Storm Emission}}.
\bjtitle{\solphys}
\bvolume{202},
\bfpage{337}.
\doiurl{10.1023/A:1012211412695}.
\adsurl{http://ads.bao.ac.cn/abs/2001SoPh..202..337C}.
\end{barticle}
\endbibitem

\bibitem[\protect\citeauthoryear{{Crooker}, {Gosling}, and
  {Kahler}}{2002}]{Crooker02}
\begin{barticle}
\bauthor{\bsnm{{Crooker}}, \binits{N.U.}},
\bauthor{\bsnm{{Gosling}}, \binits{J.T.}},
\bauthor{\bsnm{{Kahler}}, \binits{S.W.}}:
\byear{2002},
\batitle{{Reducing heliospheric magnetic flux from coronal mass ejections
  without disconnection}}.
\bjtitle{Journal of Geophysical Research (Space Physics)}
\bvolume{107},
\bfpage{1028}.
\doiurl{10.1029/2001JA000236}.
\adsurl{http://ads.bao.ac.cn/abs/2002JGRA..107.1028C}.
\end{barticle}
\endbibitem

\bibitem[\protect\citeauthoryear{{Del Zanna} \textit{et~al.}}{2011}]{Zanna11}
\begin{barticle}
\bauthor{\bsnm{{Del Zanna}}, \binits{G.}},
\bauthor{\bsnm{{Aulanier}}, \binits{G.}},
\bauthor{\bsnm{{Klein}}, \binits{K.-L.}},
\bauthor{\bsnm{{T{\"o}r{\"o}k}}, \binits{T.}}:
\byear{2011},
\batitle{{A single picture for solar coronal outflows and radio noise storms}}.
\bjtitle{\aap}
\bvolume{526},
\bfpage{A137}.
\doiurl{10.1051/0004-6361/201015231}.
\adsurl{http://ads.bao.ac.cn/abs/2011A\%26A...526A.137D}.
\end{barticle}
\endbibitem

\bibitem[\protect\citeauthoryear{{Edmondson}}{2012}]{Edmondson12}
\begin{barticle}
\bauthor{\bsnm{{Edmondson}}, \binits{J.K.}}:
\byear{2012},
\batitle{{On the Role of Interchange Reconnection in the Generation of the Slow
  Solar Wind}}.
\bjtitle{\ssr}
\bvolume{172},
\bfpage{209}.
\doiurl{10.1007/s11214-011-9767-y}.
\adsurl{http://ads.bao.ac.cn/abs/2012SSRv..172..209E}.
\end{barticle}
\endbibitem

\bibitem[\protect\citeauthoryear{{Elgar{\o}y}}{1977}]{Elgaroy77}
\begin{bbook}
\bauthor{\bsnm{{Elgar{\o}y}}, \binits{E.{\O}.}}:
\byear{1977},
\bbtitle{{Solar noise storms.}}
\adsurl{http://ads.bao.ac.cn/abs/1977sns..book.....E}.
\end{bbook}
\endbibitem

\bibitem[\protect\citeauthoryear{{Fisk}}{2005}]{Fisk05}
\begin{barticle}
\bauthor{\bsnm{{Fisk}}, \binits{L.A.}}:
\byear{2005},
\batitle{{The Open Magnetic Flux of the Sun. I. Transport by Reconnections with
  Coronal Loops}}.
\bjtitle{\apj}
\bvolume{626},
\bfpage{563}.
\doiurl{10.1086/429957}.
\adsurl{http://ads.bao.ac.cn/abs/2005ApJ...626..563F}.
\end{barticle}
\endbibitem

\bibitem[\protect\citeauthoryear{{Golub} \textit{et~al.}}{2007}]{Golub07}
\begin{barticle}
\bauthor{\bsnm{{Golub}}, \binits{L.}},
\bauthor{\bsnm{{Deluca}}, \binits{E.}},
\bauthor{\bsnm{{Austin}}, \binits{G.}},
\bauthor{\bsnm{{Bookbinder}}, \binits{J.}},
\bauthor{\bsnm{{Caldwell}}, \binits{D.}},
\bauthor{\bsnm{{Cheimets}}, \binits{P.}},
\bauthor{\bsnm{{Cirtain}}, \binits{J.}},
\bauthor{\bsnm{{Cosmo}}, \binits{M.}},
\bauthor{\bsnm{{Reid}}, \binits{P.}},
\bauthor{\bsnm{{Sette}}, \binits{A.}},
\bauthor{\bsnm{{Weber}}, \binits{M.}},
\bauthor{\bsnm{{Sakao}}, \binits{T.}},
\bauthor{\bsnm{{Kano}}, \binits{R.}},
\bauthor{\bsnm{{Shibasaki}}, \binits{K.}},
\bauthor{\bsnm{{Hara}}, \binits{H.}},
\bauthor{\bsnm{{Tsuneta}}, \binits{S.}},
\bauthor{\bsnm{{Kumagai}}, \binits{K.}},
\bauthor{\bsnm{{Tamura}}, \binits{T.}},
\bauthor{\bsnm{{Shimojo}}, \binits{M.}},
\bauthor{\bsnm{{McCracken}}, \binits{J.}},
\bauthor{\bsnm{{Carpenter}}, \binits{J.}},
\bauthor{\bsnm{{Haight}}, \binits{H.}},
\bauthor{\bsnm{{Siler}}, \binits{R.}},
\bauthor{\bsnm{{Wright}}, \binits{E.}},
\bauthor{\bsnm{{Tucker}}, \binits{J.}},
\bauthor{\bsnm{{Rutledge}}, \binits{H.}},
\bauthor{\bsnm{{Barbera}}, \binits{M.}},
\bauthor{\bsnm{{Peres}}, \binits{G.}},
\bauthor{\bsnm{{Varisco}}, \binits{S.}}:
\byear{2007},
\batitle{{The X-Ray Telescope (XRT) for the Hinode Mission}}.
\bjtitle{\solphys}
\bvolume{243},
\bfpage{63}.
\doiurl{10.1007/s11207-007-0182-1}.
\adsurl{http://ads.bao.ac.cn/abs/2007SoPh..243...63G}.
\end{barticle}
\endbibitem

\bibitem[\protect\citeauthoryear{{Hagenaar} and {Shine}}{2005}]{Hagenaar05}
\begin{barticle}
\bauthor{\bsnm{{Hagenaar}}, \binits{H.J.}},
\bauthor{\bsnm{{Shine}}, \binits{R.A.}}:
\byear{2005},
\batitle{{Moving Magnetic Features around Sunspots}}.
\bjtitle{\apj}
\bvolume{635},
\bfpage{659}.
\doiurl{10.1086/497367}.
\adsurl{http://ads.bao.ac.cn/abs/2005ApJ...635..659H}.
\end{barticle}
\endbibitem

\bibitem[\protect\citeauthoryear{{Harvey} and {Harvey}}{1973}]{Harvey73}
\begin{barticle}
\bauthor{\bsnm{{Harvey}}, \binits{K.}},
\bauthor{\bsnm{{Harvey}}, \binits{J.}}:
\byear{1973},
\batitle{{Observations of Moving Magnetic Features near Sunspots}}.
\bjtitle{\solphys}
\bvolume{28},
\bfpage{61}.
\doiurl{10.1007/BF00152912}.
\adsurl{http://ads.bao.ac.cn/abs/1973SoPh...28...61H}.
\end{barticle}
\endbibitem

\bibitem[\protect\citeauthoryear{{Hey}}{1946}]{Hey46}
\begin{barticle}
\bauthor{\bsnm{{Hey}}, \binits{J.S.}}:
\byear{1946},
\batitle{{Solar Radiations in the 4-6 Metre Radio Wave-Length Band}}.
\bjtitle{\nat}
\bvolume{157},
\bfpage{47}.
\doiurl{10.1038/157047b0}.
\adsurl{http://ads.bao.ac.cn/abs/1946Natur.157...47H}.
\end{barticle}
\endbibitem

\bibitem[\protect\citeauthoryear{{Iwai} \textit{et~al.}}{2012a}]{iwai12a}
\begin{barticle}
\bauthor{\bsnm{{Iwai}}, \binits{K.}},
\bauthor{\bsnm{{Miyoshi}}, \binits{Y.}},
\bauthor{\bsnm{{Masuda}}, \binits{S.}},
\bauthor{\bsnm{{Shimojo}}, \binits{M.}},
\bauthor{\bsnm{{Shiota}}, \binits{D.}},
\bauthor{\bsnm{{Inoue}}, \binits{S.}},
\bauthor{\bsnm{{Tsuchiya}}, \binits{F.}},
\bauthor{\bsnm{{Morioka}}, \binits{A.}},
\bauthor{\bsnm{{Misawa}}, \binits{H.}}:
\byear{2012}a,
\batitle{{Solar Radio Type-I Noise Storm Modulated by Coronal Mass Ejections}}.
\bjtitle{\apj}
\bvolume{744},
\bfpage{167}.
\doiurl{10.1088/0004-637X/744/2/167}.
\adsurl{http://ads.bao.ac.cn/abs/2012ApJ...744..167I}.
\end{barticle}
\endbibitem

\bibitem[\protect\citeauthoryear{{Iwai} \textit{et~al.}}{2012b}]{iwai12b}
\begin{bchapter}
\bauthor{\bsnm{{Iwai}}, \binits{K.}},
\bauthor{\bsnm{{Misawa}}, \binits{H.}},
\bauthor{\bsnm{{Tsuchiya}}, \binits{F.}},
\bauthor{\bsnm{{Morioka}}, \binits{A.}},
\bauthor{\bsnm{{Masuda}}, \binits{S.}},
\bauthor{\bsnm{{Miyoshi}}, \binits{Y.}}:
\byear{2012}b,
\bctitle{{Survey of Accelerated Particles in a Solar Active Region Using
  Hinode/XRT and Ground-Based Type-I Radio Burst Observations}}.
In: \beditor{\bsnm{{Sekii}}, \binits{T.}},
\beditor{\bsnm{{Watanabe}}, \binits{T.}},
\beditor{\bsnm{{Sakurai}}, \binits{T.}} (eds.)
\bbtitle{Hinode-3: The 3rd Hinode Science Meeting},
\bsertitle{Astronomical Society of the Pacific Conference Series}
\bseriesno{454},
\bfpage{249}.
\adsurl{http://ads.bao.ac.cn/abs/2012ASPC..454..249I}.
\end{bchapter}
\endbibitem

\bibitem[\protect\citeauthoryear{{Kai}, {Melrose}, and {Suzuki}}{1985}]{Kai85}
\begin{bbook}
\bauthor{\bsnm{{Kai}}, \binits{K.}},
\bauthor{\bsnm{{Melrose}}, \binits{D.B.}},
\bauthor{\bsnm{{Suzuki}}, \binits{S.}}:
\byear{1985},
In: \beditor{\bsnm{{McLean}}, \binits{D.J.}},
\beditor{\bsnm{{Labrum}}, \binits{N.R.}} (eds.)
\bbtitle{{Storms}},
\bfpage{415}.
\adsurl{http://ads.bao.ac.cn/abs/1985srph.book..415K}.
\end{bbook}
\endbibitem

\bibitem[\protect\citeauthoryear{{Kathiravan}, {Ramesh}, and
  {Nataraj}}{2007}]{Kath07}
\begin{barticle}
\bauthor{\bsnm{{Kathiravan}}, \binits{C.}},
\bauthor{\bsnm{{Ramesh}}, \binits{R.}},
\bauthor{\bsnm{{Nataraj}}, \binits{H.S.}}:
\byear{2007},
\batitle{{The Post-Coronal Mass Ejection Solar Atmosphere and Radio Noise Storm
  Activity}}.
\bjtitle{\apjl}
\bvolume{656},
\bfpage{L37}.
\doiurl{10.1086/512013}.
\adsurl{http://ads.bao.ac.cn/abs/2007ApJ...656L..37K}.
\end{barticle}
\endbibitem

\bibitem[\protect\citeauthoryear{{Kerdraon} and {Delouis}}{1997}]{Kerdraon97}
\begin{bchapter}
\bauthor{\bsnm{{Kerdraon}}, \binits{A.}},
\bauthor{\bsnm{{Delouis}}, \binits{J.-M.}}:
\byear{1997},
\bctitle{{The Nan{\c c}ay Radioheliograph}}.
In: \beditor{\bsnm{{Trottet}}, \binits{G.}} (ed.)
\bbtitle{Coronal Physics from Radio and Space Observations},
\bsertitle{Lecture Notes in Physics, Berlin Springer Verlag}
\bseriesno{483},
\bfpage{192}.
\doiurl{10.1007/BFb0106458}.
\adsurl{http://ads.bao.ac.cn/abs/1997LNP...483..192K}.
\end{bchapter}
\endbibitem

\bibitem[\protect\citeauthoryear{{Kosugi} \textit{et~al.}}{2007}]{Kosugi07}
\begin{barticle}
\bauthor{\bsnm{{Kosugi}}, \binits{T.}},
\bauthor{\bsnm{{Matsuzaki}}, \binits{K.}},
\bauthor{\bsnm{{Sakao}}, \binits{T.}},
\bauthor{\bsnm{{Shimizu}}, \binits{T.}},
\bauthor{\bsnm{{Sone}}, \binits{Y.}},
\bauthor{\bsnm{{Tachikawa}}, \binits{S.}},
\bauthor{\bsnm{{Hashimoto}}, \binits{T.}},
\bauthor{\bsnm{{Minesugi}}, \binits{K.}},
\bauthor{\bsnm{{Ohnishi}}, \binits{A.}},
\bauthor{\bsnm{{Yamada}}, \binits{T.}},
\bauthor{\bsnm{{Tsuneta}}, \binits{S.}},
\bauthor{\bsnm{{Hara}}, \binits{H.}},
\bauthor{\bsnm{{Ichimoto}}, \binits{K.}},
\bauthor{\bsnm{{Suematsu}}, \binits{Y.}},
\bauthor{\bsnm{{Shimojo}}, \binits{M.}},
\bauthor{\bsnm{{Watanabe}}, \binits{T.}},
\bauthor{\bsnm{{Shimada}}, \binits{S.}},
\bauthor{\bsnm{{Davis}}, \binits{J.M.}},
\bauthor{\bsnm{{Hill}}, \binits{L.D.}},
\bauthor{\bsnm{{Owens}}, \binits{J.K.}},
\bauthor{\bsnm{{Title}}, \binits{A.M.}},
\bauthor{\bsnm{{Culhane}}, \binits{J.L.}},
\bauthor{\bsnm{{Harra}}, \binits{L.K.}},
\bauthor{\bsnm{{Doschek}}, \binits{G.A.}},
\bauthor{\bsnm{{Golub}}, \binits{L.}}:
\byear{2007},
\batitle{{The Hinode (Solar-B) Mission: An Overview}}.
\bjtitle{\solphys}
\bvolume{243},
\bfpage{3}.
\doiurl{10.1007/s11207-007-9014-6}.
\adsurl{http://ads.bao.ac.cn/abs/2007SoPh..243....3K}.
\end{barticle}
\endbibitem

\bibitem[\protect\citeauthoryear{{Lemen} \textit{et~al.}}{2012}]{Lemen12}
\begin{barticle}
\bauthor{\bsnm{{Lemen}}, \binits{J.R.}},
\bauthor{\bsnm{{Title}}, \binits{A.M.}},
\bauthor{\bsnm{{Akin}}, \binits{D.J.}},
\bauthor{\bsnm{{Boerner}}, \binits{P.F.}},
\bauthor{\bsnm{{Chou}}, \binits{C.}},
\bauthor{\bsnm{{Drake}}, \binits{J.F.}},
\bauthor{\bsnm{{Duncan}}, \binits{D.W.}},
\bauthor{\bsnm{{Edwards}}, \binits{C.G.}},
\bauthor{\bsnm{{Friedlaender}}, \binits{F.M.}},
\bauthor{\bsnm{{Heyman}}, \binits{G.F.}},
\bauthor{\bsnm{{Hurlburt}}, \binits{N.E.}},
\bauthor{\bsnm{{Katz}}, \binits{N.L.}},
\bauthor{\bsnm{{Kushner}}, \binits{G.D.}},
\bauthor{\bsnm{{Levay}}, \binits{M.}},
\bauthor{\bsnm{{Lindgren}}, \binits{R.W.}},
\bauthor{\bsnm{{Mathur}}, \binits{D.P.}},
\bauthor{\bsnm{{McFeaters}}, \binits{E.L.}},
\bauthor{\bsnm{{Mitchell}}, \binits{S.}},
\bauthor{\bsnm{{Rehse}}, \binits{R.A.}},
\bauthor{\bsnm{{Schrijver}}, \binits{C.J.}},
\bauthor{\bsnm{{Springer}}, \binits{L.A.}},
\bauthor{\bsnm{{Stern}}, \binits{R.A.}},
\bauthor{\bsnm{{Tarbell}}, \binits{T.D.}},
\bauthor{\bsnm{{Wuelser}}, \binits{J.-P.}},
\bauthor{\bsnm{{Wolfson}}, \binits{C.J.}},
\bauthor{\bsnm{{Yanari}}, \binits{C.}},
\bauthor{\bsnm{{Bookbinder}}, \binits{J.A.}},
\bauthor{\bsnm{{Cheimets}}, \binits{P.N.}},
\bauthor{\bsnm{{Caldwell}}, \binits{D.}},
\bauthor{\bsnm{{Deluca}}, \binits{E.E.}},
\bauthor{\bsnm{{Gates}}, \binits{R.}},
\bauthor{\bsnm{{Golub}}, \binits{L.}},
\bauthor{\bsnm{{Park}}, \binits{S.}},
\bauthor{\bsnm{{Podgorski}}, \binits{W.A.}},
\bauthor{\bsnm{{Bush}}, \binits{R.I.}},
\bauthor{\bsnm{{Scherrer}}, \binits{P.H.}},
\bauthor{\bsnm{{Gummin}}, \binits{M.A.}},
\bauthor{\bsnm{{Smith}}, \binits{P.}},
\bauthor{\bsnm{{Auker}}, \binits{G.}},
\bauthor{\bsnm{{Jerram}}, \binits{P.}},
\bauthor{\bsnm{{Pool}}, \binits{P.}},
\bauthor{\bsnm{{Soufli}}, \binits{R.}},
\bauthor{\bsnm{{Windt}}, \binits{D.L.}},
\bauthor{\bsnm{{Beardsley}}, \binits{S.}},
\bauthor{\bsnm{{Clapp}}, \binits{M.}},
\bauthor{\bsnm{{Lang}}, \binits{J.}},
\bauthor{\bsnm{{Waltham}}, \binits{N.}}:
\byear{2012},
\batitle{{The Atmospheric Imaging Assembly (AIA) on the Solar Dynamics
  Observatory (SDO)}}.
\bjtitle{\solphys}
\bvolume{275},
\bfpage{17}.
\doiurl{10.1007/s11207-011-9776-8}.
\adsurl{http://ads.bao.ac.cn/abs/2012SoPh..275...17L}.
\end{barticle}
\endbibitem

\bibitem[\protect\citeauthoryear{{Mercier} \textit{et~al.}}{1984}]{Mercier84}
\begin{barticle}
\bauthor{\bsnm{{Mercier}}, \binits{C.}},
\bauthor{\bsnm{{Elgaroy}}, \binits{O.}},
\bauthor{\bsnm{{Tlamicha}}, \binits{A.}},
\bauthor{\bsnm{{Zlobec}}, \binits{P.}}:
\byear{1984},
\batitle{{Solar noise storms coordinated observations - May 16-24, 1981}}.
\bjtitle{\solphys}
\bvolume{92},
\bfpage{375}.
\doiurl{10.1007/BF00157259}.
\adsurl{http://ads.bao.ac.cn/abs/1984SoPh...92..375M}.
\end{barticle}
\endbibitem

\bibitem[\protect\citeauthoryear{{Pesnell}, {Thompson}, and
  {Chamberlin}}{2012}]{Pesnell12}
\begin{barticle}
\bauthor{\bsnm{{Pesnell}}, \binits{W.D.}},
\bauthor{\bsnm{{Thompson}}, \binits{B.J.}},
\bauthor{\bsnm{{Chamberlin}}, \binits{P.C.}}:
\byear{2012},
\batitle{{The Solar Dynamics Observatory (SDO)}}.
\bjtitle{\solphys}
\bvolume{275},
\bfpage{3}.
\doiurl{10.1007/s11207-011-9841-3}.
\adsurl{http://ads.bao.ac.cn/abs/2012SoPh..275....3P}.
\end{barticle}
\endbibitem

\bibitem[\protect\citeauthoryear{{Schatten}, {Wilcox}, and
  {Ness}}{1969}]{Schat69}
\begin{barticle}
\bauthor{\bsnm{{Schatten}}, \binits{K.H.}},
\bauthor{\bsnm{{Wilcox}}, \binits{J.M.}},
\bauthor{\bsnm{{Ness}}, \binits{N.F.}}:
\byear{1969},
\batitle{{A model of interplanetary and coronal magnetic fields}}.
\bjtitle{\solphys}
\bvolume{6},
\bfpage{442}.
\doiurl{10.1007/BF00146478}.
\adsurl{http://ads.bao.ac.cn/abs/1969SoPh....6..442S}.
\end{barticle}
\endbibitem

\bibitem[\protect\citeauthoryear{{Schou} \textit{et~al.}}{2012}]{Schou12}
\begin{barticle}
\bauthor{\bsnm{{Schou}}, \binits{J.}},
\bauthor{\bsnm{{Scherrer}}, \binits{P.H.}},
\bauthor{\bsnm{{Bush}}, \binits{R.I.}},
\bauthor{\bsnm{{Wachter}}, \binits{R.}},
\bauthor{\bsnm{{Couvidat}}, \binits{S.}},
\bauthor{\bsnm{{Rabello-Soares}}, \binits{M.C.}},
\bauthor{\bsnm{{Bogart}}, \binits{R.S.}},
\bauthor{\bsnm{{Hoeksema}}, \binits{J.T.}},
\bauthor{\bsnm{{Liu}}, \binits{Y.}},
\bauthor{\bsnm{{Duvall}}, \binits{T.L.}},
\bauthor{\bsnm{{Akin}}, \binits{D.J.}},
\bauthor{\bsnm{{Allard}}, \binits{B.A.}},
\bauthor{\bsnm{{Miles}}, \binits{J.W.}},
\bauthor{\bsnm{{Rairden}}, \binits{R.}},
\bauthor{\bsnm{{Shine}}, \binits{R.A.}},
\bauthor{\bsnm{{Tarbell}}, \binits{T.D.}},
\bauthor{\bsnm{{Title}}, \binits{A.M.}},
\bauthor{\bsnm{{Wolfson}}, \binits{C.J.}},
\bauthor{\bsnm{{Elmore}}, \binits{D.F.}},
\bauthor{\bsnm{{Norton}}, \binits{A.A.}},
\bauthor{\bsnm{{Tomczyk}}, \binits{S.}}:
\byear{2012},
\batitle{{Design and Ground Calibration of the Helioseismic and Magnetic Imager
  (HMI) Instrument on the Solar Dynamics Observatory (SDO)}}.
\bjtitle{\solphys}
\bvolume{275},
\bfpage{229}.
\doiurl{10.1007/s11207-011-9842-2}.
\adsurl{http://ads.bao.ac.cn/abs/2012SoPh..275..229S}.
\end{barticle}
\endbibitem

\bibitem[\protect\citeauthoryear{{Schrijver} and {De Rosa}}{2003}]{Schri03}
\begin{barticle}
\bauthor{\bsnm{{Schrijver}}, \binits{C.J.}},
\bauthor{\bsnm{{De Rosa}}, \binits{M.L.}}:
\byear{2003},
\batitle{{Photospheric and heliospheric magnetic fields}}.
\bjtitle{\solphys}
\bvolume{212},
\bfpage{165}.
\doiurl{10.1023/A:1022908504100}.
\adsurl{http://ads.bao.ac.cn/abs/2003SoPh..212..165S}.
\end{barticle}
\endbibitem

\bibitem[\protect\citeauthoryear{{Sheeley}}{1969}]{Sheeley69}
\begin{barticle}
\bauthor{\bsnm{{Sheeley}}, \binits{N.R.} \bsuffix{Jr.}}:
\byear{1969},
\batitle{{The Evolution of the Photospheric Network}}.
\bjtitle{\solphys}
\bvolume{9},
\bfpage{347}.
\doiurl{10.1007/BF02391657}.
\adsurl{http://ads.bao.ac.cn/abs/1969SoPh....9..347S}.
\end{barticle}
\endbibitem

\bibitem[\protect\citeauthoryear{{Shimizu}}{1994}]{Shimizu94}
\begin{botherref}
\oauthor{\bsnm{{Shimizu}}, \binits{T.}}:
1994,
{in T. Sakurai, T. Hirayama and G. Ai (eds.)}.
\textit{Proc. of the 2nd Japan-China Seminar on Solar Phys.}
\textbf{193}.
\end{botherref}
\endbibitem

\bibitem[\protect\citeauthoryear{{Spicer}, {Benz}, and {Huba}}{1982}]{Spicer82}
\begin{barticle}
\bauthor{\bsnm{{Spicer}}, \binits{D.S.}},
\bauthor{\bsnm{{Benz}}, \binits{A.O.}},
\bauthor{\bsnm{{Huba}}, \binits{J.D.}}:
\byear{1982},
\batitle{{Solar type I noise storms and newly emerging magnetic flux}}.
\bjtitle{\aap}
\bvolume{105},
\bfpage{221}.
\adsurl{http://ads.bao.ac.cn/abs/1982A\%26A...105..221S}.
\end{barticle}
\endbibitem

\bibitem[\protect\citeauthoryear{{Willson}}{2005a}]{Willson05b}
\begin{barticle}
\bauthor{\bsnm{{Willson}}, \binits{R.F.}}:
\byear{2005}a,
\batitle{{Collaborative VLA, SOHO and RHESSI observations of evolving sources
  of energy release in the corona above active regions}}.
\bjtitle{Advances in Space Research}
\bvolume{35},
\bfpage{1813}.
\doiurl{10.1016/j.asr.2005.04.081}.
\adsurl{http://ads.bao.ac.cn/abs/2005AdSpR..35.1813W}.
\end{barticle}
\endbibitem

\bibitem[\protect\citeauthoryear{{Willson}}{2005b}]{Willson05a}
\begin{barticle}
\bauthor{\bsnm{{Willson}}, \binits{R.F.}}:
\byear{2005}b,
\batitle{{Very Large Array and SOHO Observations of Type I Noise Storms,
  Large-Scale Loops and Magnetic Restructuring in the Corona}}.
\bjtitle{\solphys}
\bvolume{227},
\bfpage{311}.
\doiurl{10.1007/s11207-005-1104-8}.
\adsurl{http://ads.bao.ac.cn/abs/2005SoPh..227..311W}.
\end{barticle}
\endbibitem

\bibitem[\protect\citeauthoryear{{Zwaan}}{1978}]{Zwaan78}
\begin{barticle}
\bauthor{\bsnm{{Zwaan}}, \binits{C.}}:
\byear{1978},
\batitle{{On the Appearance of Magnetic Flux in the Solar Photosphere}}.
\bjtitle{\solphys}
\bvolume{60},
\bfpage{213}.
\doiurl{10.1007/BF00156523}.
\adsurl{http://ads.bao.ac.cn/abs/1978SoPh...60..213Z}.
\end{barticle}
\endbibitem

\bibitem[\protect\citeauthoryear{{Zwaan}}{1985}]{Zwaan85}
\begin{barticle}
\bauthor{\bsnm{{Zwaan}}, \binits{C.}}:
\byear{1985},
\batitle{{The emergence of magnetic flux}}.
\bjtitle{\solphys}
\bvolume{100},
\bfpage{397}.
\doiurl{10.1007/BF00158438}.
\adsurl{http://ads.bao.ac.cn/abs/1985SoPh..100..397Z}.
\end{barticle}
\endbibitem

\end{thebibliography}

     % Checking: look if the file containing the ``\bibitem'' exits
     %           so check if the .bbl file exist (bibTeX compilation)
\IfFileExists{\jobname.bbl}{} {\typeout{}
\typeout{****************************************************}
\typeout{****************************************************}
\typeout{** Please run "bibtex \jobname" to obtain} \typeout{**
the bibliography and then re-run LaTeX} \typeout{** twice to fix
the references !}
\typeout{****************************************************}
\typeout{****************************************************}
\typeout{}}

%\begin{figure*}[!ht]
%\centering
%\includegraphics[trim=2.5cm 1cm 2.5cm 0.5cm,scale=0.7,angle=90]{figs/AIASi.eps}
%\caption{Left: Cut-off AIA 171 \A image with a outlined box region, which is the IRIS raster image field-of-view of 141\arcsec $\times$ %175\arcsec. Right: IRIS \Si raster image.}
%\label{figaiasi}
%\end{figure*}
%\newpage
 \begin{figure*}[!ht]    %%%%%%%%%%%%%%%%%% FIGURE 1
   \centerline{\includegraphics[trim=2.5cm 3.5cm 13cm 3cm,scale=0.6]{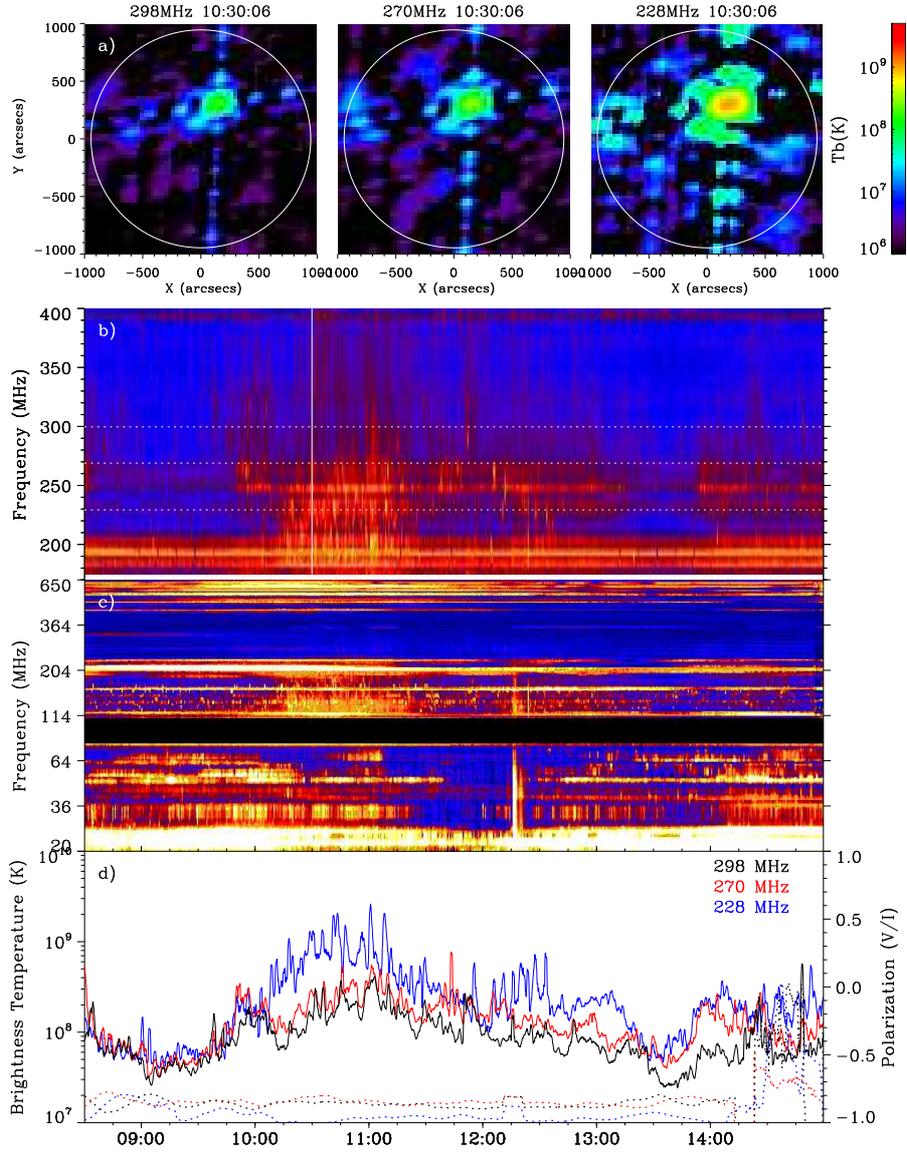}
              }
              \caption{The type-I radio bursts observed on 30 July 2011. (a) The NRH radio images at 298, 270, and 228 MHz, (b) the spectral data given by a composition of data from the CALLISTO-BLEN and (c) ARTEMIS IV radio spectrometers, and (d) the $T_\mathrm{B}$ (solid) and the polarization level (dashed) observed by NRH at the three frequencies 298 (black), 270 (red), and 228 (blue) MHz. The polarization is calculated as the ratio of the NRH Stokes V over Stokes I. The vertical white line in (b) denotes the time of the NRH images. A movie is available online.}
   \label{f1}
   \end{figure*}
\begin{figure*}[!ht]    %%%%%%%%%%%%%%%%%% FIGURE 2
   \centerline{\includegraphics[trim=0cm 6cm 0cm 5cm,scale=0.6]{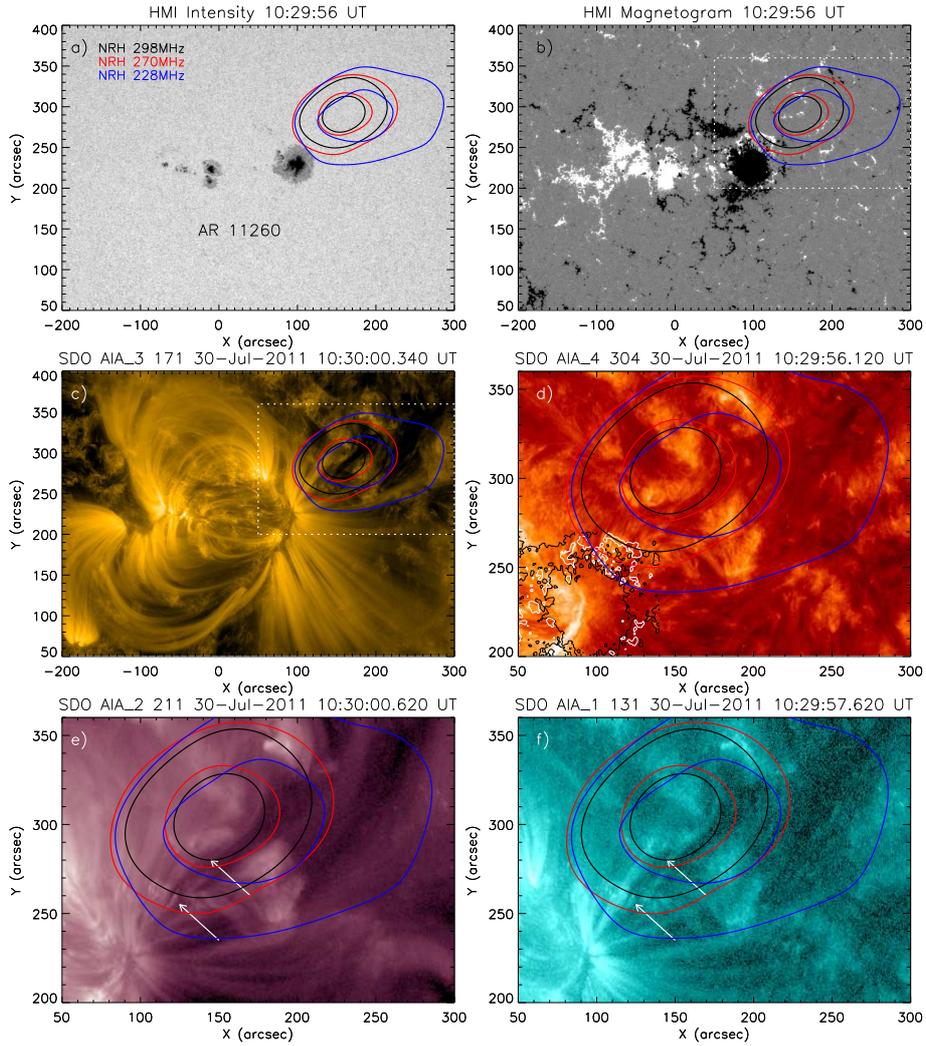}
              }
              \caption{The SDO HMI and AIA data with superposed NRH contours of the radio sources. The contours represent levels at 70\,\% and 90\,\% of the corresponding $T_\mathrm{B}$ maximum, at 298 (black), 270 (red), and 228 (blue) MHz. The $\pm$60 G　 contours of the line-of-sight component of the HMI magnetic field ($B_\mathrm{LOS}$) are overlaid on panel d. The FOV of panels d\,--\,f is shown as the dashed square in panels b and c. The white arrows in panels e\,--\,f point to the lower (just above the MMFs) and higher EUV activities that are suggested to be correlated with the type-I radio burst. Two accompanying movies are available online. One shows the temporal evolution of EUV structures and magnetic fields of the entire AR (11260) as well as the accompanying radio sources, the other presents the localized region with the brightening structures and relevant perturbations at various EUV wavelengths.}
   \label{f2}
   \end{figure*}
%%%%%%%%%%%%%%%%%%%%%%%%%%
\begin{figure*}[!ht]    %%%%%%%%%%%%%%%%%% FIGURE 3
   \centerline{\includegraphics[trim=0.cm 0.2cm 0cm 0.5cm,scale=0.9]{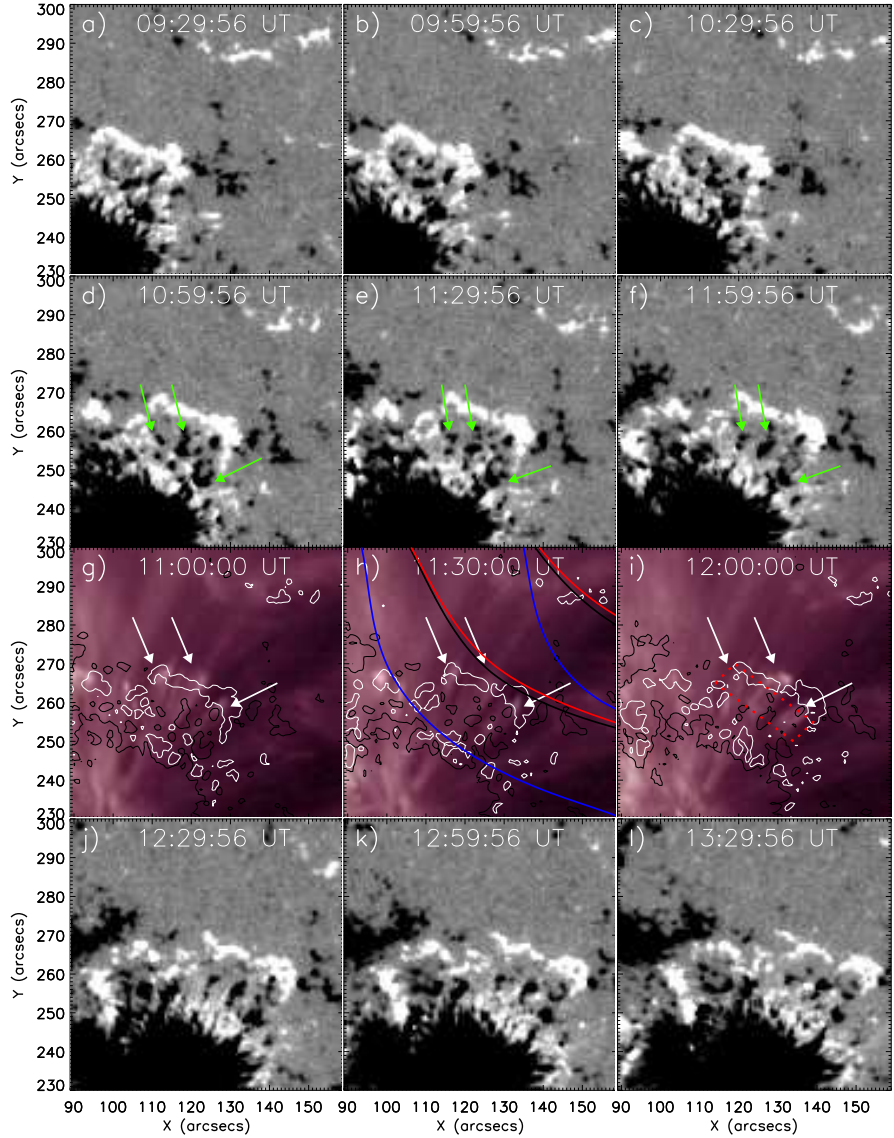}
              }
              \caption{The evolutionary sequence of the photospheric magnetic field given by HMI/SDO from 09:30 to 13:30 UT on 30 July 2011. The green arrows point to the three sets of negative magnetic polarities within the regime of interest. Panels g\,--\,i show the coronal activity observed by AIA at 211 \AA. The white arrows point to the three brightening stripes. Overlaid on panels g\,--\,i are the $\pm$60 G　 contours of the line-of-sight component of the HMI magnetic field ($B_\mathrm{LOS}$) and the NRH contours at 90\,\% and 70\,\% of the corresponding $T_\mathrm{B}$ maximum at 298 (black), 270 (red), and 228 (blue) MHz. The red dashed square in panel i denotes the slice used to obtain the distance-time map of Figure 4. The slice is 29\arcsec\ long and 12\arcsec\ wide. A movie is available online, which reveals the evolution of the magnetic field on 30 July 2011. The red rectangle in the movie corresponds to the FOV of Figure 3.}
   \label{f3}
   \end{figure*}
%%%%%%%%%%%%%%%%%%%%%%%%%%
\begin{figure*}[!ht]    %%%%%%%%%%%%%%%%%% FIGURE 4
   \centerline{\includegraphics[trim=1.5cm 0cm 1.cm 0cm,scale=0.55,angle=90]{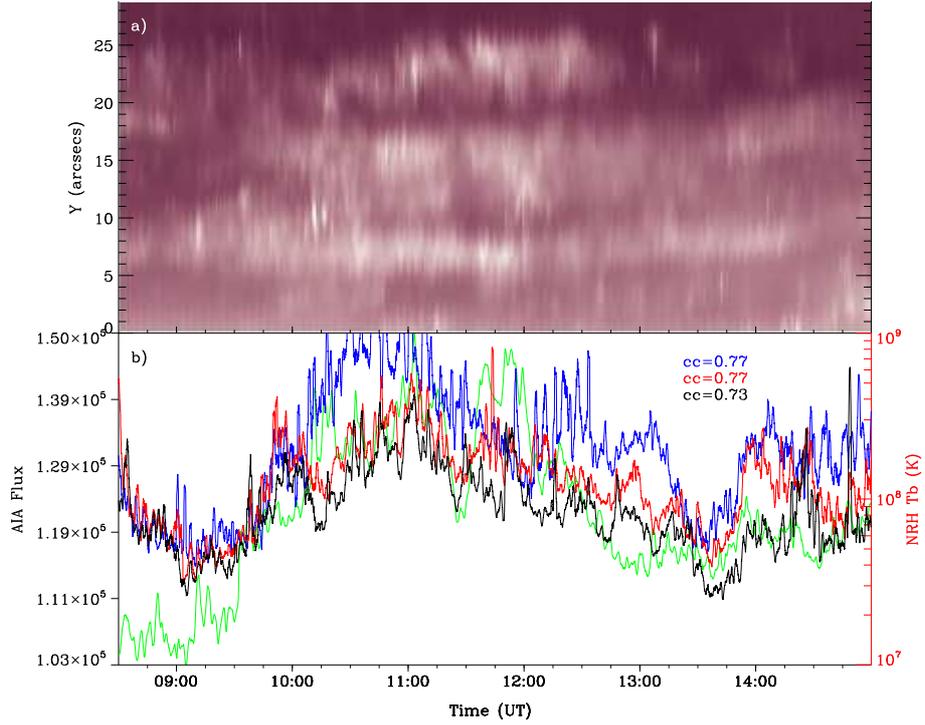}
              }
              \caption{(a) Distance-time map at 211 \AA　 along the slice presented in Figure 3i. (b) Temporal profiles of EUV 211 \AA　 emission intensity (green) and the $T_\mathrm{B}$ maxima profiles obtained by NRH at 298 (black), 270 (red), and 228 (blue) MHz.}
   \label{f4}
   \end{figure*}
%%%%%%%%%%%%%%%%%%%%%%%%%%
\begin{figure*}[!ht]    %%%%%%%%%%%%%%%%%% FIGURE 5
   \centerline{\includegraphics[trim= 1.0cm 1.2cm -4.5cm 1cm,scale=0.9,angle=90]{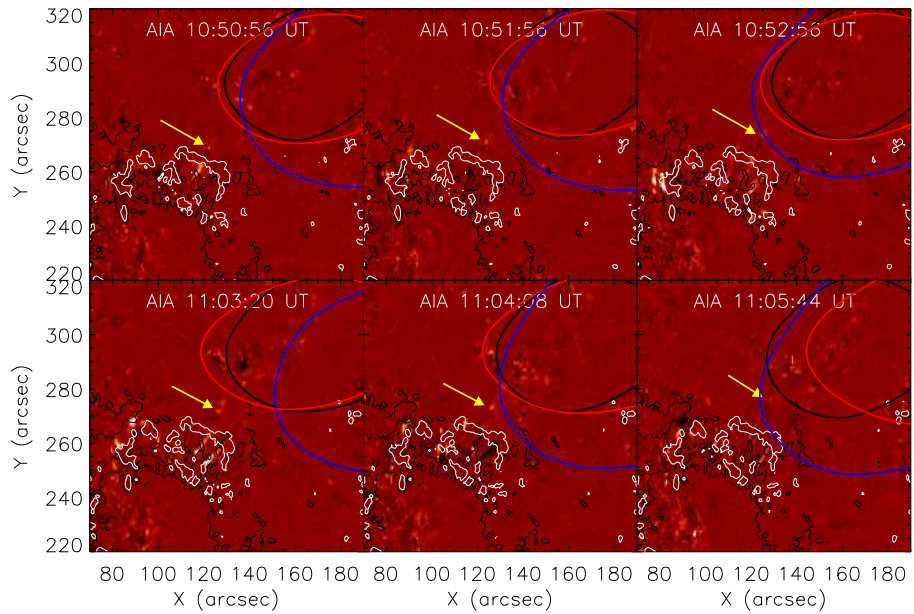}
              }
              \caption{Two examples of bi-directional plasma flows observed at 304 \AA, marked by yellow arrows. Overlaid on the panels are the $\pm$60 G 　contours of the line-of-sight component of the HMI magnetic field ($B_\mathrm{LOS}$) and the NRH contours at 90\,\% of the corresponding $T_\mathrm{B}$ maximum at 298 (black), 270 (red), and 228 (blue) MHz. A movie is available online.}
   \label{f5}
   \end{figure*}
%%%%%%%%%%%%%%%%%%%%%%%%%%
\begin{figure*}[!ht]    %%%%%%%%%%%%%%%%%% FIGURE 6
   \centerline{\includegraphics[trim=0.0cm 0.3cm 0.cm 0cm,scale=0.45]{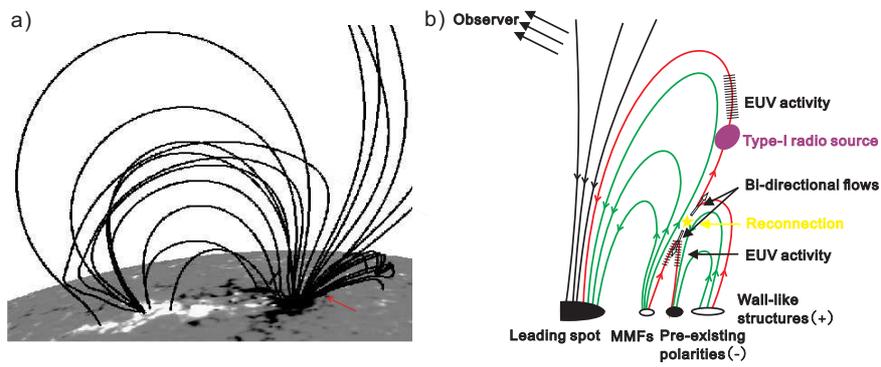}
              }
              \caption{(a) Large-scale coronal magnetic field configuration of AR 11260 obtained from the PFSS extrapolation model. The photospheric magnetic field is from the Carrington maps (CR 2112\,--\,2113). The red arrow points to the northwest neighborhood of the leading spot. (b) Schematic of the ambient magnetic structures and the generation of the type-I burst. White and black regions at the bottom (the photosphere) mean positive and negative magnetic polarity. The wall-like structures, the pre-existing polarities, the MMFs, and the leading spot constitute a quadrupolar configuration. The green lines denote the magnetic configuration in which the reconnection occurs (see the yellow star). The red field lines indicate the post-reconnection loops. The purple region indicates the type-I noise storm source. The fences represent EUV activity at lower and higher altitudes.
              }
   \label{f6}
   \end{figure*}

\end{article}

\end{document}